\documentclass[%
amsmath,amssymb,
prl,
twocolumn
]{revtex4-1}

\usepackage{graphicx}
\usepackage{dcolumn}
\usepackage{bm}
\usepackage{xcolor}
\usepackage[normalem]{ulem}
\usepackage{amsmath}
\usepackage{algorithm}
\usepackage[noend]{algpseudocode}

\begin{document}

\title{Exposing the hidden influence of selection rules on phonon-phonon scattering by pressure and temperature tuning}

\author{Navaneetha K. Ravichandran}
\email{navaneeth@iisc.ac.in}
\affiliation{Department of Mechanical Engineering, Indian Institute of Science, Bangalore, Karnataka 560012, India}

\author{David Broido}
\affiliation{Department of Physics, Boston College, Chestnut Hill, MA 02467, USA}


\begin{abstract}
Using \emph{ab initio} calculations, we show that the hidden influence of selection rules on three-phonon scattering can be exposed through anomalous signatures in the pressure ($P$) and temperature ($T$) dependence of the thermal conductivities, $\kappa$, of certain compounds.  Boron phosphide reveals such underlying behavior through an exceptionally sharp initial rise in $\kappa$ with increasing $P$, which may be the steepest of any material, and also a peak and decrease in $\kappa$ at high $P$. These features are in stark contrast to the measured behavior for many solids, and occur at experimentally accessible conditions. Similar anomalous behavior is predicted for silicon carbide and other related materials.
\end{abstract}
\maketitle
The thermal conductivity, $\kappa$, is a fundamental transport parameter that governs the efficiency of heat conduction through solids.  In insulating crystals, the heat is transported by phonons, and intrinsic thermal resistance comes from phonon-phonon scattering processes, which arise from the anharmonicity of the interatomic bonding potential~\cite{peierls_zur_1929, peierls_quantum_1955, ziman_electrons_2001}. Lowest-order processes involving the mutual interaction among three phonons typically dominate this intrinsic resistance.\\\indent
For a given phonon mode, the collection of all such scattering processes that conserve energy and momentum - \emph{the scattering phase space} - is dictated entirely by a material's phonon dispersions.  Specific features in these dispersions activate selection rules that can severely restrict the phase space of certain phonon-phonon scattering channels~\cite{peierls_zur_1929, peierls_quantum_1955, ziman_electrons_2001, herpin_contribution_1952, klemens_thermal_1958, klemens_decay_1967, orbach_attenuation_1964, lax_spontaneous_1981, lindsay_first-principles_2013, broido_ab_2013, mukhopadhyay_optic_2016, ravichandran_phonon-phonon_2020}. While several of these selection rules have recently been identified~\cite{lindsay_first-principles_2013, broido_ab_2013, mukhopadhyay_optic_2016, ravichandran_phonon-phonon_2020}, the most well-known of them dictates that an acoustic phonon cannot decay anharmonically into a set of others that are in the same dispersive phonon branch~\cite{peierls_zur_1929, peierls_quantum_1955, ziman_electrons_2001, herpin_contribution_1952, klemens_thermal_1958, klemens_decay_1967, orbach_attenuation_1964, lax_spontaneous_1981}.  It explains the long lifetimes of large wave vector transverse acoustic (TA) phonons, observed at low temperatures~\cite{grill_excitation_1975, ulbrich_propagation_1980}. Exceptions exist~\cite{tamura_spontaneous_1985}, but the phase space for those scattering events is small~\cite{fNote1}.\\\indent
The influence of this selection rule on $\kappa$ was long thought to be negligible since inter-branch phonon scattering processes dominate thermal transport, particularly at higher temperatures. However, it was recently pointed out that, even around and beyond room temperature (RT; 300 K), this selection rule can be activated for high frequency phonons in materials where the three acoustic phonon branches come close to each other in some region of the Brillouin zone (BZ), thereby mimicking the behavior of phonon decay within a single branch~\cite{lindsay_first-principles_2013, broido_ab_2013, ravichandran_phonon-phonon_2020}.  The phase space available to a heat carrying, high frequency acoustic (\emph{A}) phonon in such a region of the BZ, decaying into two others (\emph{AAA} process) can then become quite small, thus resulting in its long lifetime and large contributions to $\kappa$.\\\indent
Perhaps the most interesting example of the impact of this \emph{AAA} selection rule is cubic boron arsenide (BAs).  First principles calculations~\cite{lindsay_first-principles_2013} predicted that unusually weak \emph{AAA} scattering, resulting from activation of the \emph{AAA} selection rule, contributes to a $\kappa$ for BAs that should be comparable to that of diamond, the crystal having the highest measured $\kappa$ of all materials at RT. While inclusion of higher-order four-phonon scattering was found to reduce the BAs $\kappa$ ~\cite{feng_four-phonon_2017, tian_unusual_2018}, it still achieved a predicted RT $\kappa$ value of around 1300 Wm$^{-1}$K$^{-1}$, by far the highest value of any naturally occurring material, behind only the carbon crystals (e.g. diamond and graphite). These theoretical findings have been confirmed by measurements~\cite{tian_unusual_2018, kang_experimental_2018, li_high_2018}, adding strong support to the impact of selection rules and corresponding reductions in the three-phonon scattering phase space on $\kappa$.\\\indent
Recently, we have identified several other compounds, such as cubic boron phosphide (BP) and silicon carbide (SiC), where the \emph{AAA} selection rule is activated, giving anomalously weak \emph{AAA} scattering channels~\cite{ravichandran_phonon-phonon_2020}. However, unlike in BAs, for these compounds, the weak \emph{AAA} scattering is masked by another strong three-phonon scattering channel involving two acoustic phonons and an optic (\emph{O}) phonon (\emph{AAO} scattering channel), thus drastically reducing its impact on $\kappa$.  In this letter, we show that the otherwise-hidden weak \emph{AAA} scattering in BP and SiC can be exposed by hydrostatic pressure ($P$) and temperature ($T$) tuning of $\kappa$. We demonstrate that the evolving interplay between \emph{AAA} and \emph{AAO} scattering channels with varying $P$ and $T$ gives rise to two remarkable features in the $\kappa$ of BP around and above RT - an unusually sharp rise in $\kappa$ as $P$ increases from ambient pressure and, a peak and subsequent drop in $\kappa$ at high $P$.  These features are contrary to the measured $\kappa$($P$, $T$) of many other insulating crystals that instead show a roughly linear increase with $P$ far away from phase transitions~\cite{hakansson_thermal_1986, gerlich_temperature_1982, hakansson_thermal_1986, dalton_effect_2013, ohta_lattice_2012, manthilake_lattice_2011, hofmeister_pressure_2007, hofmeister_mantle_1999, hakansson_thermal_1986, andersson_thermal_1985}, consistent with simple theories~\cite{hofmeister_pressure_2007, slack_thermal_1979}.\\\indent
The $\kappa$($P$, $T$) of BP is calculated using a recently developed unified \emph{ab initio} theoretical framework, which combines the use of density functional theory to obtain phonon dispersions and phonon scattering rates, with a numerical solution of the phonon Boltzmann transport equation~\cite{ravichandran_unified_2018}. The theory has no adjustable parameters, and it has demonstrated good agreement with the measured $\kappa$ of several materials over a broad temperature and pressure range~\cite{ravichandran_unified_2018, tian_unusual_2018, ravichandran_non-monotonic_2019, chen_ultrahigh_2020, ravichandran_phonon-phonon_2020}, including BP and SiC.  Both three-phonon and higher-order four-phonon scattering processes are included in the calculations along with phonon scattering by mass disorder arising from the natural isotope mixture on the constituent atoms~\cite{fNote2}. The details of this first principles approach have been published in Ref.~\cite{ravichandran_unified_2018} and are summarized in the Supplemental Material (SM) section S1~\cite{fNote3}. We present the results for BP here, while those for SiC can be found in the SM section S4~\cite{fNote3}.\\\indent
\begin{figure}
\begin{center}
\includegraphics[scale=0.45, trim=7mm 10mm 0mm 0mm, clip]{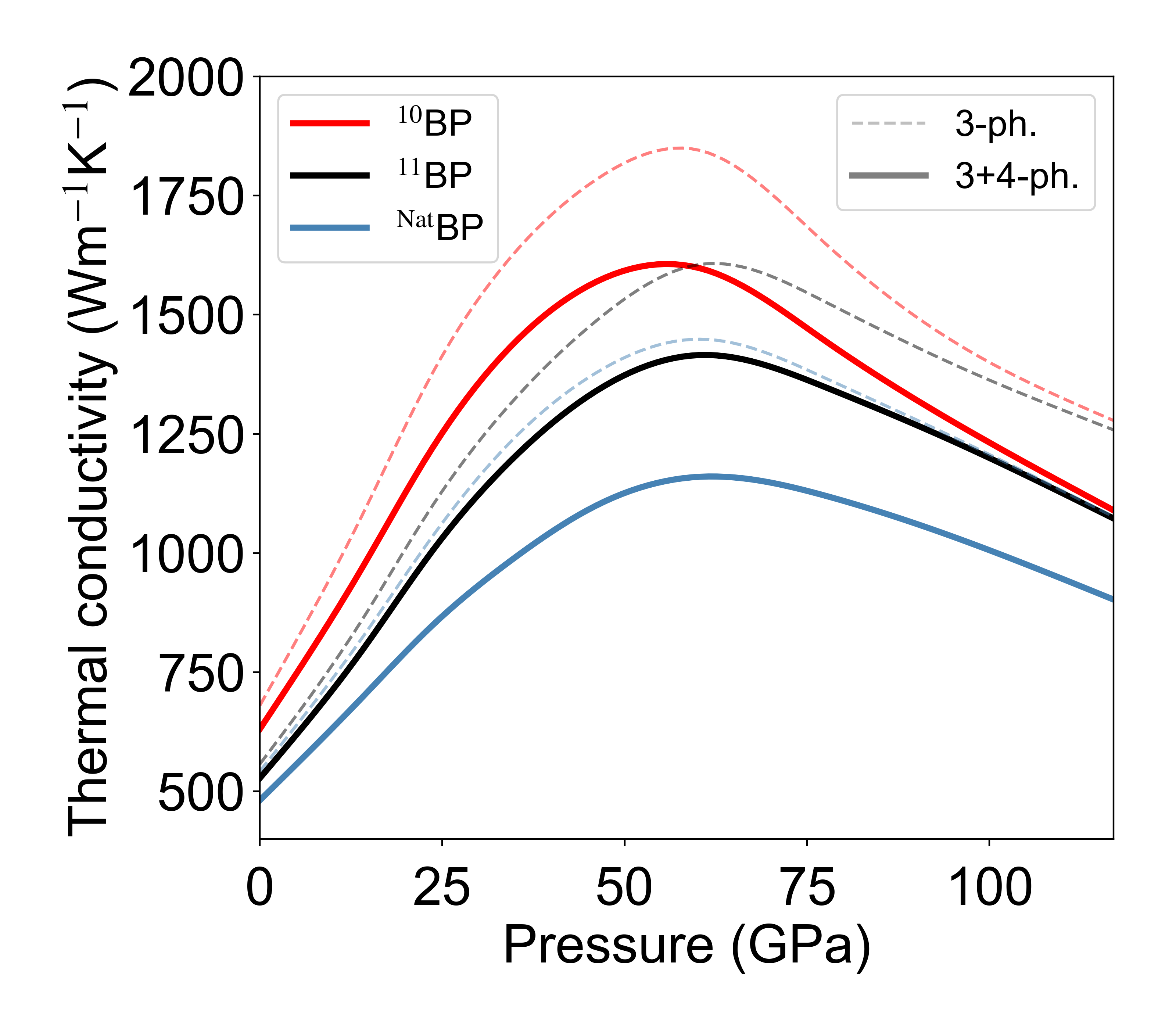}
\end{center}
\caption{RT $\kappa$ of BP with naturally occurring B isotope concentration, isotopically pure $^{11}$B, and isotopically pure $^{10}$B as a function of $P$, with (solid) and without (dashed) four-phonon scattering. All curves show a peak and drop in $\kappa$ with increasing $P$, unlike many other materials.}	\label{fig_1}
\end{figure}
Figure~\ref{fig_1} shows the calculated RT $\kappa$($P$) of BP, with and without four-phonon scattering, for three cases:  BP with (1) naturally occurring B isotope mix ($^{\mathrm{Nat}}$BP), (2) B atoms isotopically enriched to 100 \% $^{11}$B ($^{11}$BP) and (3) B atoms isotopically enriched to 100 \% $^{10}$B ($^{10}$BP). All curves peak at around 55-60 GPa and subsequently decrease with increasing $P$, contrary to the typically measured linearly increasing behavior for many other materials. In addition, the $\kappa$($P$) curves rise rapidly with increasing $P$, and there is a significant increase in the slopes, $\mathrm{d}\kappa/\mathrm{d}P$, going from $^{\mathrm{Nat}}$BP to $^{11}$BP to $^{10}$BP.  The peak values of $\kappa$ are extremely large, around 1150 Wm$^{-1}$K$^{-1}$ ($^{\mathrm{Nat}}$BP), 1400 Wm$^{-1}$K$^{-1}$ ($^{11}$BP) and 1600 Wm$^{-1}$K$^{-1}$ ($^{10}$BP), about 2.5 times higher than their values at ambient pressure. It is important to note that, the effect of four-phonon scattering on $\kappa$ is relatively weak at all pressures in BP. Thus, the non-monotonic $\kappa$($P$) behavior for BP is unrelated to that predicted for BAs~\cite{ravichandran_non-monotonic_2019}, as discussed below.\\\indent
To understand these pressure-dependencies of $\kappa$, we first examine the phonon-phonon scattering processes in BP at ambient pressure.  The light atoms and stiff bonding of BP result in a high $\kappa$ of around 500-600 Wm$^{-1}$K$^{-1}$ at RT and ambient pressure~\cite{zheng_high_2018, chen_ultrahigh_2020, dames_ultrahigh_2018}. BP crystallizes in the zinc blende structure and has two atoms in each unit cell, so the phonon dispersions consist of three \emph{A} and three \emph{O} phonon branches. Only three types of three-phonon scattering processes can occur, which involve combinations of \emph{A} and \emph{O} phonons: \emph{AAA}, \emph{AAO}, and \emph{AOO}.  Energy conservation forbids all other processes (e.g. \emph{OOO})~\cite{ravichandran_phonon-phonon_2020}.  Figure~\ref{fig_2} (b) shows the RT \emph{AAA}, \emph{AAO} and \emph{AOO} scattering rates for BP at various pressures, along with those for four-phonon scattering. The sharp dip in the \emph{AAA} scattering rates at $P$=0 in Fig.~\ref{fig_2} (b)  results from the \emph{AAA} selection rule driven by the close proximity of the three acoustic phonon branches in a small region of the BZ, as seen in Fig.~\ref{fig_2} (a), along $\Gamma\to K$ direction.  Similar dips driven by the same selection rule occur in other cubic compounds such as BAs, BSb, SiC, diamond, c-BN, and GaN~\cite{broido_ab_2013, ravichandran_phonon-phonon_2020} and certain transition metal carbides~\cite{li_fermi_2018}. Some examples are shown in Fig. S1 in the SM.\\
\begin{figure*}[!ht]
\begin{center}
\includegraphics[scale=0.31, trim=2mm 2mm 0mm 0mm, clip]{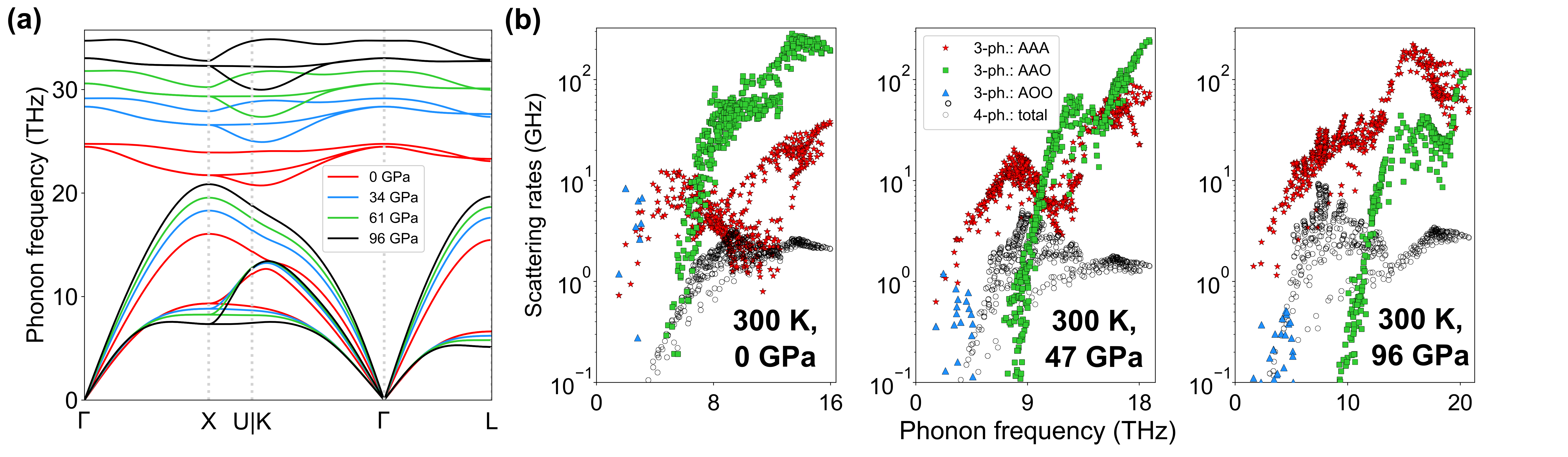}
\end{center}
\caption{\textbf{(a)} Phonon dispersions of $^{11}$BP at different values of $P$. \textbf{(b)} Process-wise three-phonon and total four-phonon scattering rates for $^{11}$BP at RT and different pressures.  Since acoustic phonons carry the majority of the heat, only their frequency bandwidth is plotted.}	\label{fig_2}
\end{figure*}
Apart from the close proximity of the acoustic phonons, other features in the phonon dispersions of BP and SiC activate additional selection rules. The mass difference between constituent atoms of these compound semiconductors causes a frequency gap, $\Delta\omega_{A-O}$, between acoustic and optic phonons (A-O gap).  Energy conservation mandates that only those acoustic phonons with frequencies greater than $\Delta\omega_{A-O}$ can participate in \emph{AAO} scattering processes~\cite{lindsay_first-principles_2013}. Similarly, energy conservation restricts the participation of acoustic phonons in \emph{AOO} scattering processes to those with frequencies less than the optic phonon bandwidth, $\Delta\omega_O$~\cite{mukhopadhyay_optic_2016, ravichandran_phonon-phonon_2020}. These restrictions are illustrated in Fig. S2 in the SM.\\\indent
As shown in Fig.~\ref{fig_2} (b), the \emph{AAO} scattering rates in BP dominate over almost the entire frequency region of the dip in \emph{AAA} scattering rates. The relatively small mass difference between boron and phosphorus atoms ($M_P/M_B=2.6$) gives a $\Delta\omega_{A-O}$ that is not large enough to freeze out the \emph{AAO} scattering channel for acoustic phonons above $\sim$6 THz.  This behavior is in stark contrast to large mass ratio compounds like BAs ($M_{As}/M_B=7$),  where $\Delta\omega_{A-O}$ is large enough to almost completely remove \emph{AAO} processes for acoustic phonons, thereby fully exposing the sharp dip in the \emph{AAA} scattering rates, and resulting in significantly enhanced three-phonon limited $\kappa$~\cite{lindsay_first-principles_2013, fNote4}. To illustrate the importance of the \emph{AAO} processes in BP, we perform calculations by artificially turning them off, and find that the RT $\kappa$ of $^{11}$BP at ambient pressure jumps from 530 Wm$^{-1}$K$^{-1}$ (560 Wm$^{-1}$K$^{-1}$) to over 2600 Wm$^{-1}$K$^{-1}$ (6000 Wm$^{-1}$K$^{-1}$), including three-phonon and four-phonon (three-phonon only) scattering.  These values are about twice the corresponding values in BAs.\\\indent
In BP, application of hydrostatic pressure introduces three intertwined behaviors that affect $\kappa$:  (i) the longitudinal acoustic (LA) and optic phonons shift to higher frequencies, which increases LA phonon group velocities and weakens the \emph{AAO} scattering rates due to decreased optic phonon occupations.  Both of these changes contribute to the increasing $\kappa$ with $P$ found in many materials.  In BP, two additional features are critical to explain the anomalous behavior seen in Fig.~\ref{fig_1} - (ii) optic phonons stiffen faster than do acoustic phonons [see Fig.~\ref{fig_2} (a)], which increases $\Delta\omega_{A-O}$ and shifts the onset of \emph{AAO} processes to higher frequencies, thereby exposing increasingly large portions of the \emph{AAA} dip. This behavior significantly increases the rate of rise in the $\kappa (P)$ of BP in the low $P$ range, as discussed below; and (iii) with increasing $P$, the LA phonons stiffen, while the transverse acoustic (TA) phonons weakly soften. The resulting increased separation between LA and TA phonon branches gradually removes the impact of the \emph{AAA} selection rule, and so, increases the \emph{AAA} scattering rates, relative to those at ambient pressure [see Fig.~\ref{fig_2} (b)]. This behavior acts to drive $\kappa$ lower, as pointed out previously~\cite{lindsay_anomalous_2015, ravichandran_non-monotonic_2019}.\\\indent
\emph{Peak and decrease in $\kappa$} - At RT and low $P$, the influence of trends (i) and (ii) is stronger than trend (iii) resulting in an increasing $\kappa (P)$. Beyond 50 GPa, the continued shift of \emph{AAO} processes towards higher phonon frequencies almost fully exposes the \emph{AAA} dip, and the rising \emph{AAA} scattering rates from trend (iii) eventually dominate the behavior causing $\kappa$ to decrease with increasing $P$. As a result of this evolving interplay between \emph{AAA} and \emph{AAO} processes, the RT $\kappa (P)$ achieves a peak value at around 60 GPa, which is around 2.5 times that of its value at $P=0$.\\\indent
Figure~\ref{fig_3} (a) shows the $\kappa (P)$ for $^{11}$BP scaled by its zero pressure value, $\kappa_0$, for different $T$. Around and above RT, the curves for each $T$ roughly overlap, indicating that the $P$ and $T$ dependences are separable. Such separability is also predicted from empirical theory~\cite{manthilake_lattice_2011, hofmeister_pressure_2007} and is found in conventionally behaving compounds such as MgO (see Fig. S4 in the SM). That it also occurs in BP with its anomalous non-monotonic $\kappa (P)$, is a consequence of the independent changes in \emph{AAA} and \emph{AAO} scattering rates induced by changing $T$ or $P$.  Increasing $T$ strengthens \emph{AAO} processes relative to \emph{AAA} processes, while increasing $P$ shifts the \emph{AAO} processes to higher frequencies relative to \emph{AAA} processes.  Above RT, \emph{AAO} scattering rates in BP dominate in magnitude over \emph{AAA} scattering rates until very high $P$ [Fig.~\ref{fig_3} (b)], just as they do at RT [Fig.~\ref{fig_2} (b)]. Thus, the evolution of $\kappa (P)$ above RT is qualitatively the same as that at RT.  However, for $T$ well below RT, as \emph{AAO} scattering rates weaken relative to \emph{AAA} scattering rates [see Fig.~\ref{fig_3} (b)], the rise in $\kappa$ from trends (i) and (ii) is overcome by trend (iii) at lower $P$, resulting in a weaker $P$ dependence to $\kappa (P)$, which eventually decreases below $\kappa_0$, as shown in Fig.~\ref{fig_3} (a) for $T$ = 100 K.  Similar behavior of $\kappa (P)$ also occurs in SiC (see SM section S4).\\
\begin{figure*}[!ht]
\begin{center}
\includegraphics*[scale=0.35, trim=5mm 2mm 0mm 0mm, clip]{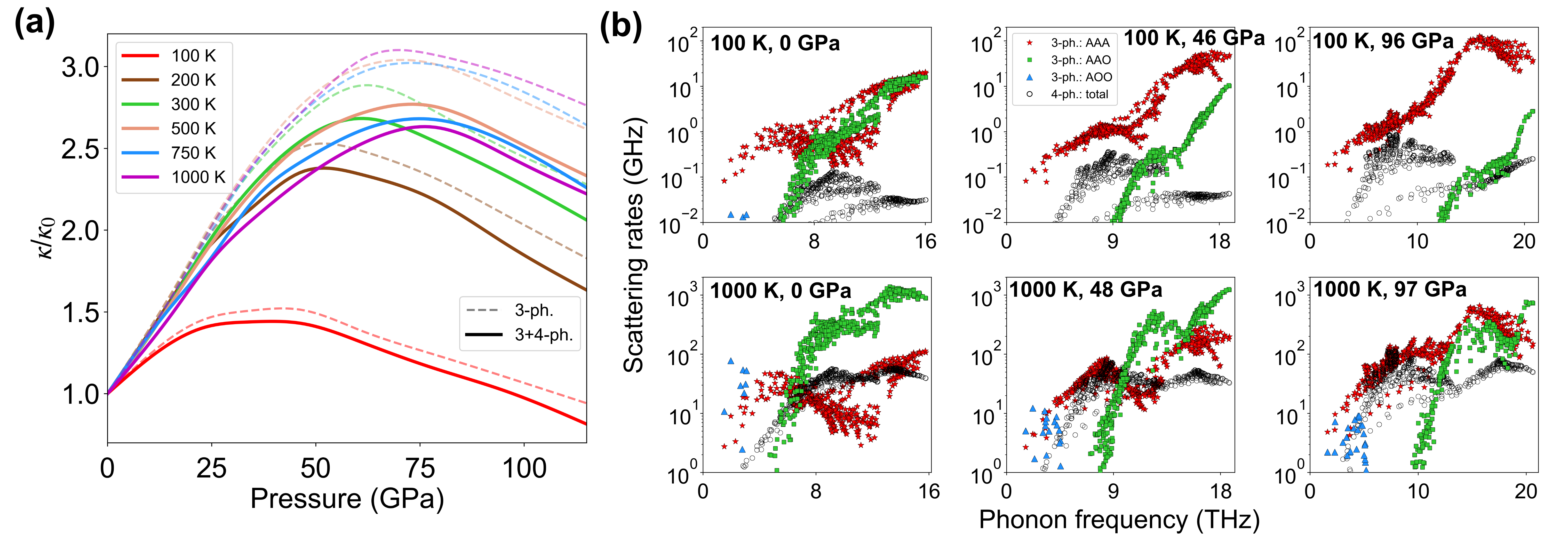}
\end{center}
\caption{\textbf{(a)} Pressure dependence of the $\kappa$ of $^{11}$BP scaled by its zero pressure values ($\kappa_0$) for different temperatures. \textbf{(b)} Process-wise three-phonon and total four-phonon scattering rates for $^{11}$BP at 100 K and 1000 K, and at different pressures.}	\label{fig_3}
\end{figure*}
Previously, we predicted a non-monotonic pressure dependence of $\kappa$ for BAs~\cite{ravichandran_non-monotonic_2019}, which was connected to an evolving competition between three-phonon and four-phonon scattering processes with increasing $P$. This competition is responsible for a significant suppression of the measured BAs $\kappa$-values compared with the lowest-order three-phonon prediction~\cite{feng_four-phonon_2017, tian_unusual_2018}. In stark contrast, however, the non-monotonic behavior of $\kappa (P)$ in BP arises from fundamentally different physics involving competition only among three-phonon scattering channels. This competition contributes to a significantly larger peak-$\kappa$ value in BP than in BAs (2.5x in BP vs. 1.1x in BAs at RT), and a much sharper increase in the BP $\kappa$ at low $P$ (discussed next), thus making it more accessible for experimental validation. We note that the effect of four-phonon scattering on the $\kappa (P)$ of BP and SiC is dominated by the increasing strength of the \emph{AAAA} scattering channel with increasing $P$ (see SM sections S3 and S4), but it is generally weak compared with the dominant three-phonon scattering channels, and so does not cause qualitative differences in the observed $\kappa (P)$ trends [see Figs.~\ref{fig_1} and~\ref{fig_3} (a)].\\\indent
\emph{Unusually sharp rise in $\kappa (P)$ above P=0} - The calculated slopes, $\mathrm{d}\kappa /\mathrm{d} P$, of the $\kappa (P)$ curves in Fig.~\ref{fig_1} around ambient pressure are exceptionally large.  Significant contributions to these large values come from the rapid increase in the intrinsic lifetimes of acoustic phonons whose frequencies lie just below the onset of the \emph{AAO} scattering.  With increasing $P$, this onset shifts to higher frequencies, exposing more of the \emph{AAA} dip. Phonons in the newly exposed region of the \emph{AAA} dip have reduced scattering rates and so give correspondingly larger contributions to $\kappa$.  This behavior is illustrated in Fig.~\ref{fig_4}, which shows, $\kappa (\nu)$, the spectral contributions to $\kappa$ as a function of phonon frequency, $\nu$, for $^{10}$BP and $^{11}$BP at different pressures and RT.  The effect is enhanced in $^{10}$BP compared with $^{11}$BP because the lighter $^{10}$B atom shifts the onset of \emph{AAO} scattering processes to higher frequencies even at $P=0$ (Fig. S3). This difference helps explain the much larger calculated RT and ambient pressure $\kappa$ of $^{10}$BP (630 Wm$^{-1}$K$^{-1}$) compared with that for $^{11}$BP (530 Wm$^{-1}$K$^{-1}$), consistent with measured results~\cite{zheng_high_2018, chen_ultrahigh_2020}.\\\indent
\begin{figure}[!ht]
\begin{center}
\includegraphics*[scale=0.45, trim=7mm 10mm 0mm 0mm, clip]{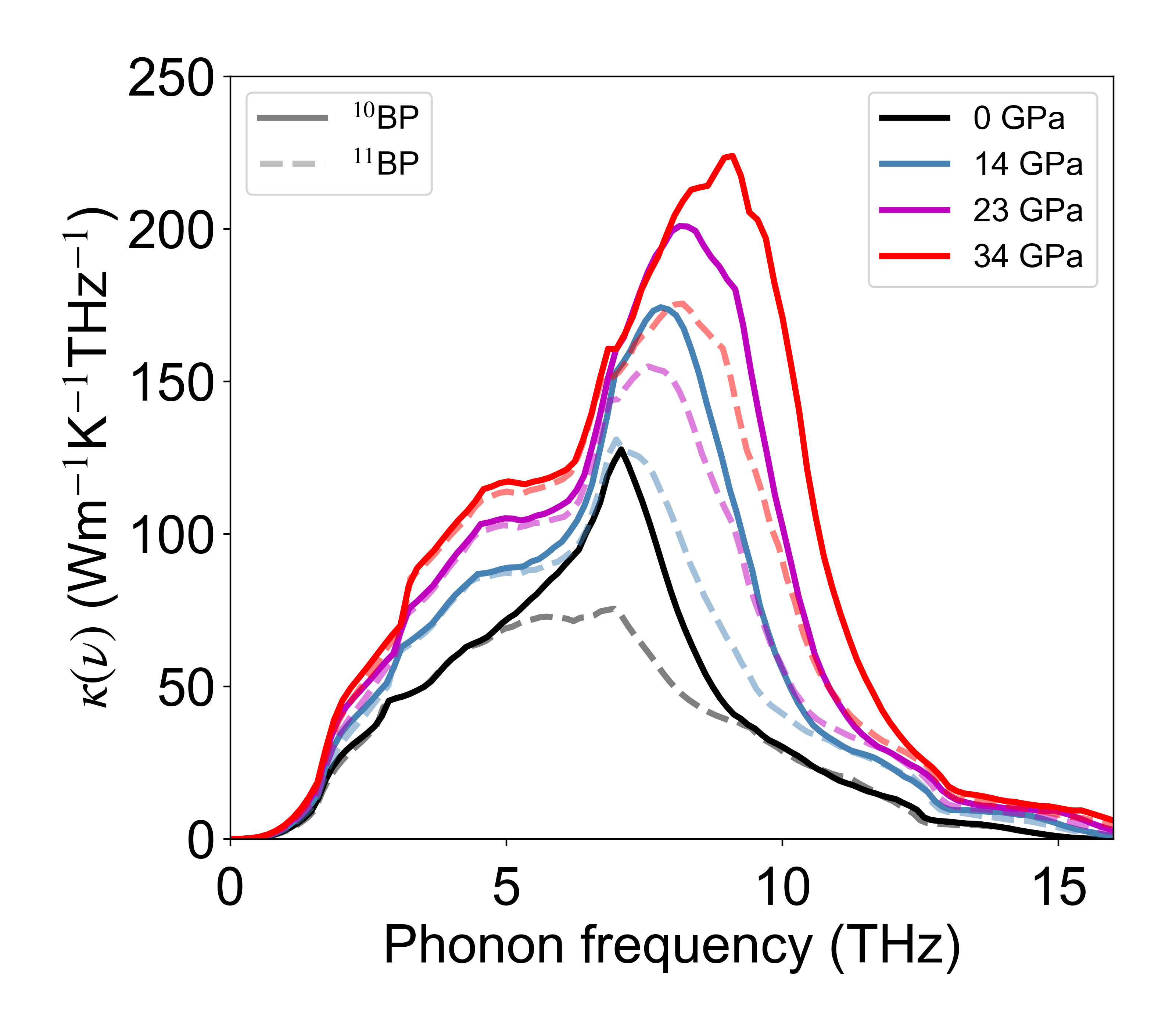}
\end{center}
\caption{Spectral contributions, $\kappa (\nu)$, to $\kappa$ of $^{10}$BP and $^{11}$BP as a function of phonon frequency, $\nu$, at different values of $P$ at RT. The sharply peaked contributions to $\kappa$ result from the upward shift (in frequency) of \emph{AAO} processes that increasingly expose the dip in the \emph{AAA} scattering rates.}	\label{fig_4}
\end{figure}
We calculate the RT $\mathrm{d}\kappa /\mathrm{d} P$ values of 15.4 Wm$^{-1}$K$^{-1}$GPa$^{-1}$ ($^{\mathrm{Nat}}$BP), 19.1 Wm$^{-1}$K$^{-1}$GPa$^{-1}$ ($^{11}$BP) and 24.1 Wm$^{-1}$K$^{-1}$GPa$^{-1}$ ($^{10}$BP) at low $P$.  Measurements and calculations of $\mathrm{d}\kappa /\mathrm{d} P$ given in the literature are only for materials with much lower $\kappa$, such as those of interest for geophysical studies~\cite{dalton_effect_2013, ohta_lattice_2012, manthilake_lattice_2011, hofmeister_mantle_1999, hofmeister_pressure_2007}, e.g. MgO, as well as alkali halides~\cite{andersson_thermal_1985, hakansson_thermal_1986, gerlich_temperature_1982}. For these materials, $\mathrm{d}\kappa /\mathrm{d} P$ is less than 3 Wm$^{-1}$K$^{-1}$GPa$^{-1}$, far smaller than our calculated values for BP.  Therefore, to put the BP values in better context, we have calculated the RT $\mathrm{d}\kappa /\mathrm{d} P$ for diamond, whose $\kappa$ is higher than that of BP. We find RT $\mathrm{d}\kappa /\mathrm{d} P$ values of 21 Wm$^{-1}$K$^{-1}$GPa$^{-1}$ and 17 Wm$^{-1}$K$^{-1}$GPa$^{-1}$ for isotopically pure and naturally occurring diamond respectively.  Remarkably, the calculated $\mathrm{d}\kappa /\mathrm{d} P$ value for $^{10}$BP is higher than that for diamond, even though the absolute RT $\kappa$ is five times lower. The present findings suggest that the $\mathrm{d}\kappa /\mathrm{d} P$ for $^{10}$BP may be the highest value achievable for any material, a striking signature of the influence of the \emph{AAA} selection rule.\\\indent
In summary, through \emph{ab initio} calculations for BP, we have shown that a hidden weak three-phonon scattering channel arising from a selection rule on anharmonic decay of acoustic phonons can be revealed through examination of the pressure dependence of the thermal conductivity, which displays anomalous features including an exceptionally fast rise and a peak with increasing pressure. We predict that such anomalous behavior will also occur in other materials similarly affected by the same selection rule, which we have demonstrated for cubic silicon carbide (SiC) in the SM (see Figs. S6-S9).  Other recently identified phase space selection rules should also give rise to unusual pressure dependence of phonon lifetimes and $\kappa$ in the affected materials~\cite{ravichandran_phonon-phonon_2020}. The present work gives new insights into the impact of selection rules on phonon-phonon scattering and demonstrates novel behavior of phonon thermal transport in solids under pressure.\\\indent
This work was supported by the Office of Naval Research under a Multidisciplinary University Research Initiative, Grant No. N00014-16-1-2436.

\bibliographystyle{unsrt}

\begin{thebibliography}{10}
\bibitem{fNote1}
We have verified the small phase space from numerical calculations for many
  compounds.
\end{thebibliography}
\onecolumngrid

\section{Supplemental Material}
\subsection{First principles theoretical approach}
We solve the linearized Boltzmann transport equation for phonons for the non-equilibrium distribution function, $n_\lambda = n^0_\lambda + n^1_\lambda$, resulting from a small applied temperature gradient, $\nabla T$, where $n^0_\lambda$ is the Bose distribution function, and $n^1_\lambda$ is the small deviation from $n^0_\lambda$ generated by $\nabla T$. Here, $\lambda\sim\left(j, \mathbf{q}\right)$ designates the phonon mode, where $j$ is the phonon branch and $\mathbf{q}$ is the phonon wave vector. The linearized Boltzmann equation is:
\begin{align}
\mathbf{v}_\lambda\cdot\nabla T\frac{\partial n^0_\lambda}{\partial T} = \frac{\partial n_\lambda}{\partial t}\Bigg|_{\mathrm{collisions}}	\label{SI:PBE}
\end{align}
where $\mathbf{v}_\lambda$ is the phonon velocity of the mode $\lambda$, and the right-hand side is the collision term, which includes three-phonon scattering, four-phonon scattering and phonon scattering by the isotopic mass disorder. Harmonic interatomic force constants (IFCs), and anharmonic third- and fourth-order IFCs are determined using the Density Functional Theory (DFT) as implemented in Quantum Espresso.  Expressing $n^1_\lambda = n^0_\lambda\left(n^0_\lambda + 1\right)\mathbf{F}_\lambda\cdot\left(-\nabla T\right)$ allows Eq.~\ref{SI:PBE} to be recast into an integral equation for the function, $\mathbf{F}_\lambda$, which is solved numerically. The thermal conductivity is then obtained as:
\begin{align}
\kappa_{\alpha\beta} = \frac{k_B}{V}\sum_\lambda\frac{\partial n^0_\lambda}{\partial T}v_{\lambda, \alpha}F_{\lambda, \alpha}
\end{align}
where $k_B$ is the Boltzmann constant, $V$ is the crystal volume, and $\kappa_{\alpha\beta}$ is the thermal conductivity tensor for heat current flow along the Cartesian direction, $\alpha$, resulting from a temperature gradient along the direction, $\beta$. For the cubic structures considered in the present work, the thermal conductivity tensor is diagonal: $\kappa_{\alpha\beta} = k\delta_{\alpha\beta}$. Further details of the computational approach including computation scheme for IFCs, expressions for the phonon scattering rates and implementation of the solution of the phonon Boltzmann equation are given in Ref.~\cite{ravichandran_unified_2018}. All the calculations in this study are performed using norm-conserving pseudopotentials with the local density approximation (LDA) for the exchange correlation. The converged parameters used in the first principles calculations in this work for BP and SiC, such as the energy cutoffs for the DFT calculations, $\mathbf{q}$-grids used in the solution of the Boltzmann equation and the cutoffs for the harmonic, cubic and quartic IFCs, can be found in Ref.~\cite{ravichandran_phonon-phonon_2020}. The pressure in our calculations is calculated by obtaining the derivative of the Helmholtz free energy, correct to fourth-order in anharmonicity ($F_{4^{\mathrm{th}}-\mathrm{order}}$), with respect to crystal volume ($V$) at each temperature ($T$) using the expression: $P(a) = -\frac{\partial F_{4^{\mathrm{th}}-\mathrm{order}}}{\partial V}\Bigg|_{T, a} \approx -\frac{F_{4^{\mathrm{th}}}(a + \Delta a) - F_{4^{\mathrm{th}}}(a - \Delta a)}{V(a + \Delta a) - V(a - \Delta a)}\Bigg|_T$, where $a$ is the crystal lattice constant and $\Delta a \sim 0.05\%$ of $a$. The complete expression for the Helmholtz free energy, correct to fourth-order in anharmonicity of the crystal potential, can be found in  section 10 of the supplemental materials in Ref.~\cite{ravichandran_non-monotonic_2019}.
\subsection{Supplemental figures to the main text for BP}
\begin{figure*}[!ht]
\begin{center}
\includegraphics*[scale=0.325]{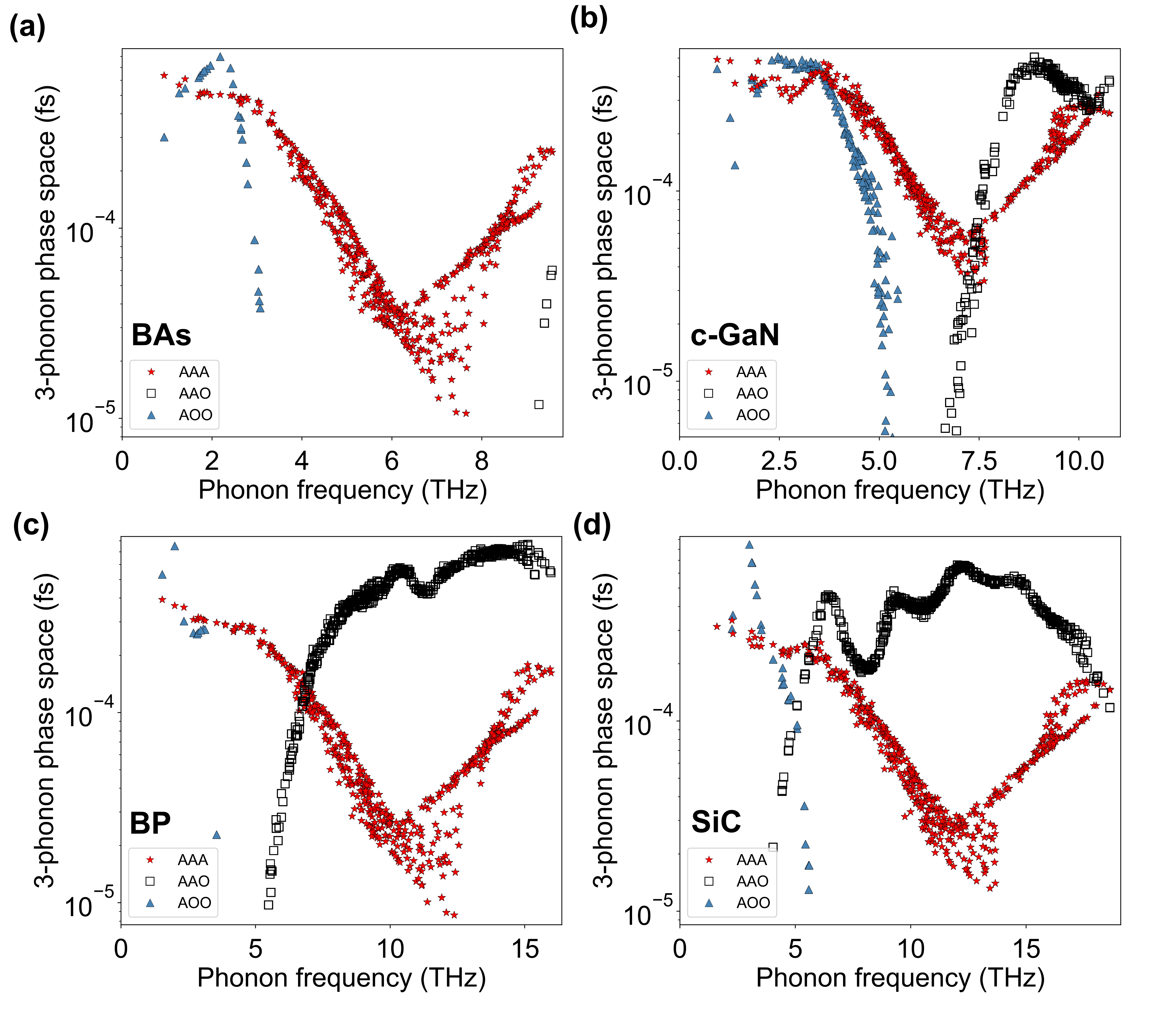}
\end{center}
\caption{Process-wise three-phonon phase space of the acoustic phonons calculated for boron arsenide [BAs - \textbf{(a)}], cubic gallium arsenide [c-GaN - \textbf{(b)}], boron phosphide [BP - \textbf{(c)}] and silicon carbide [SiC - \textbf{(d)}], organized according to decreasing mass ratio of the heavy to light constituent atoms, which correlates with a decreasing size of the frequency gap between acoustic and optic phonons (A-O gap).  The four compounds, BAs, c-GaN, BP and SiC show similar sharp dips  in the phase space for scattering between three acoustic phonons (\emph{AAA} scattering, red stars).  The large A-O gap in BAs almost completely removes the \emph{AAO} scattering (black squares) and fully exposes the dip in the \emph{AAA} scattering phase space.  In contrast, \emph{AAO} scattering processes increasingly dominate in the frequency region of the \emph{AAA} dips in c-GaN, BP and SiC as the size of the A-O gap decreases.}	\label{SI:fig_1}
\end{figure*}
\begin{figure*}[!ht]
\begin{center}
\includegraphics*[scale=0.4]{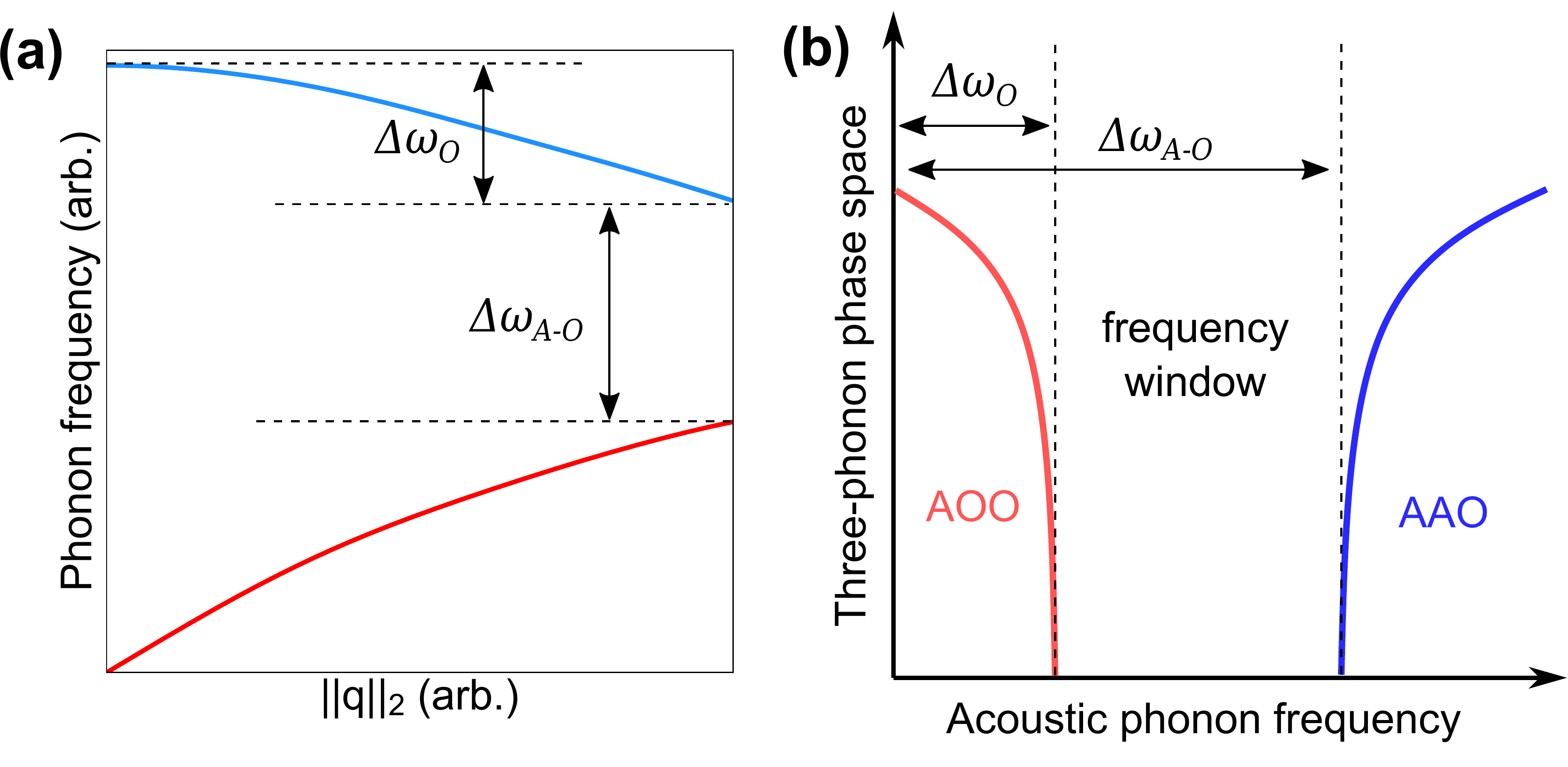}
\end{center}
\caption{\textbf{(a)} Schematic of phonon dispersions with a frequency gap between acoustic and optic phonons, $\Delta\omega_{A-O}$, and an optic phonon bandwidth of $\Delta\omega_O$. \textbf{(b)} Schematic of the corresponding three-phonon phase space for \emph{AAO} and \emph{AOO} processes illustrating the upper and lower cut-offs at $\Delta\omega_O$ and $\Delta\omega_{A-O}$, respectively.  If $\Delta\omega_{A-O} > \Delta\omega_O$, a frequency window opens in which only \emph{AAA} scattering can occur.}	\label{SI:fig_2}
\end{figure*}
\begin{figure*}[!ht]
\begin{center}
\includegraphics*[scale=0.285]{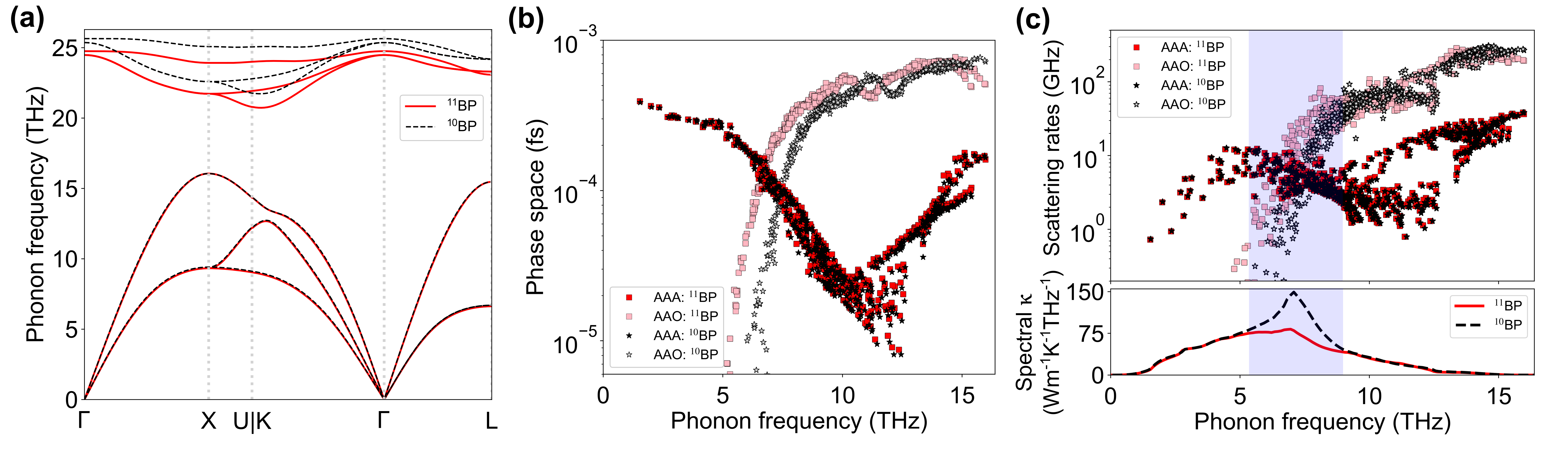}
\end{center}
\caption{\textbf{(a)} Phonon dispersions of $^{10}$BP and $^{11}$BP at ambient pressure. \textbf{(b)} Phase space for \emph{AAA} and \emph{AAO} three-phonon scattering processes for $^{10}$BP and $^{11}$BP. \textbf{(c)} Scattering rates for \emph{AAA} and \emph{AAO} three-phonon processes (top panel) and spectral contributions to $\kappa$ [$\kappa (\nu)$] (bottom panel) for $^{10}$BP and $^{11}$BP. The slightly stiffer optic phonons in $^{10}$BP compared to $^{11}$BP shifts \emph{AAO} processes to higher frequencies, thereby exposing more of the \emph{AAA} phase space dip, causing weaker \emph{AAO} scattering rates and, as a result, higher $\kappa$ contributions from a narrow frequency region (shown by the blue shaded region in \textbf{(c)}) in $^{10}$BP.}	\label{SI:fig_3}
\end{figure*}
\begin{figure*}[!ht]
\begin{center}
\includegraphics*[scale=0.5]{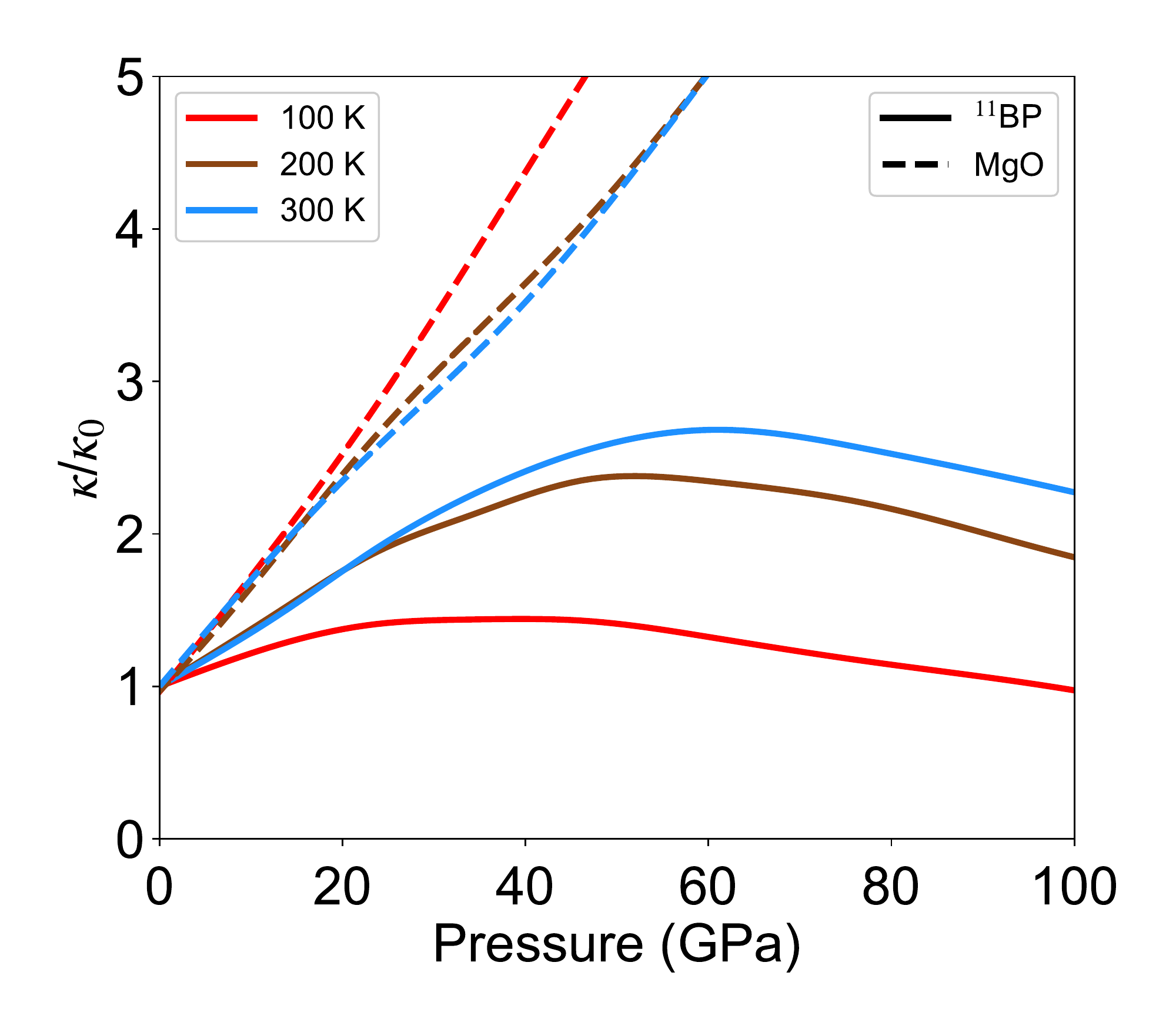}
\end{center}
\caption{Calculated $\kappa (P)$ of $^{11}$BP and MgO, scaled by their zero pressure values, $\kappa_0$, at temperatures of 100 K, 200 K and 300 K.  With decreasing temperatures, $\kappa (P)/\kappa_0$ decreases in $^{11}$BP, in striking contrast to the behavior in MgO which shows an increase in $\kappa (P)/\kappa_0$.}	\label{SI:fig_4}
\end{figure*}
\subsection{Increasing strength of four-phonon scattering in BP with increasing pressure}
\begin{figure*}[!ht]
\begin{center}
\includegraphics*[scale=0.4]{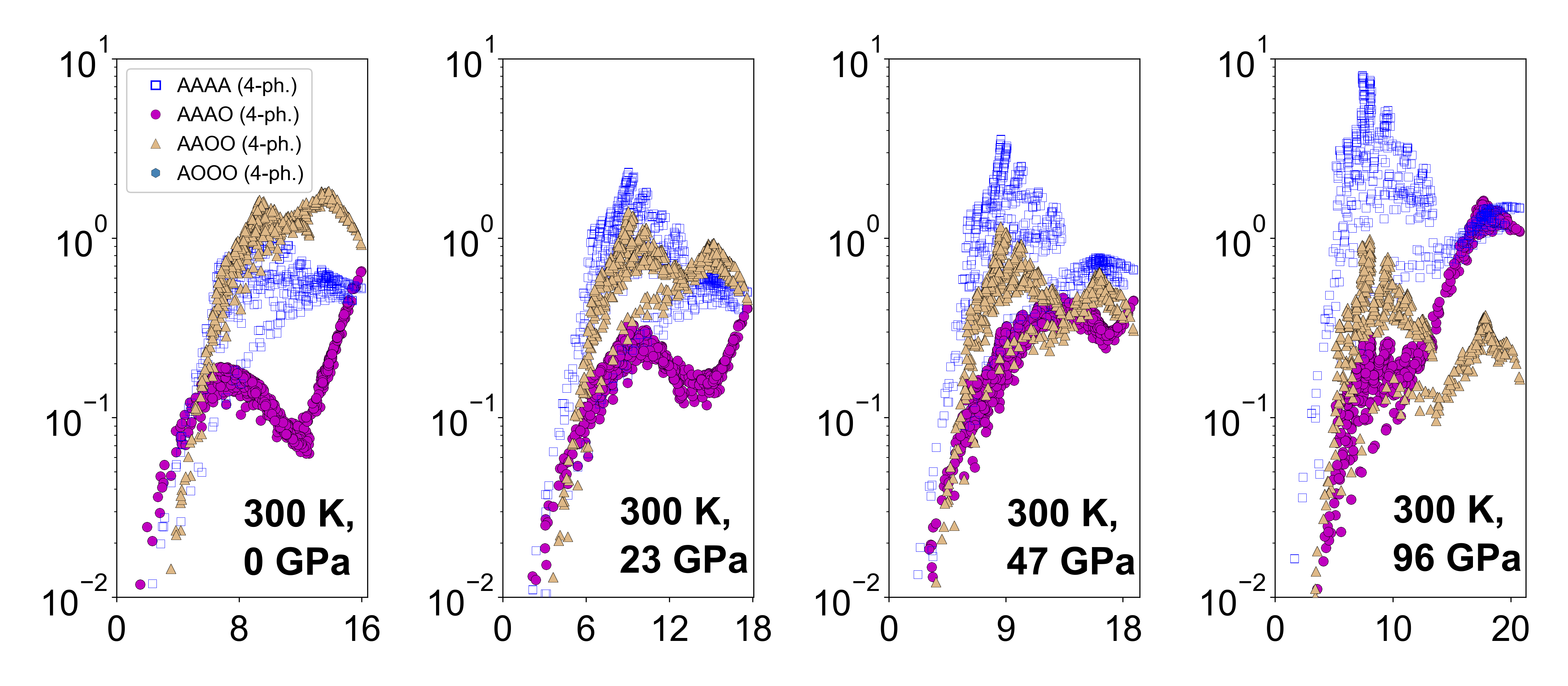}
\end{center}
\caption{Process-wise four-phonon scattering rates for BP at $T$ = 300 K at different pressures.  \emph{AAAA} scattering rates increase with increasing $P$ and dominate at high $P$. }		\label{SI:fig_5}
\end{figure*}

Figs. 1 and 3(a) in the main text show that, as $P$ increases, inclusion of four-phonon scattering increasingly suppresses the $\kappa (P)$ of BP, and that, this suppression becomes even larger at higher $T$.  This behavior is driven by strengthening of \emph{AAAA} four-phonon scattering with increasing $P$, as shown in Fig.~\ref{SI:fig_5}.  Hydrostatic pressure stiffens the chemical bonding, resulting in increased interatomic force constants (IFCs).  These increases in the fourth-order IFCs produce corresponding increases in the four-phonon scattering matrix elements acting to increase the scattering rates of all four-phonon channels.  At the same time, the increases in second-order IFCs shift optic phonons to higher frequencies, thereby reducing their populations and reducing the scattering rates of all four-phonon processes involving optic phonons, such as \emph{AAAO}, \emph{AAOO} and \emph{AOOO} processes.  As a result, \emph{AAAA} four-phonon scattering rates are increased preferentially over all other four-phonon scattering rates, which contain at least one optic phonon. \\
\pagebreak
\newpage
\clearpage
\subsection{Hidden influence of the selection rules on the $\kappa$ of SiC}
\begin{figure*}[!ht]
\begin{center}
\includegraphics*[scale=0.255]{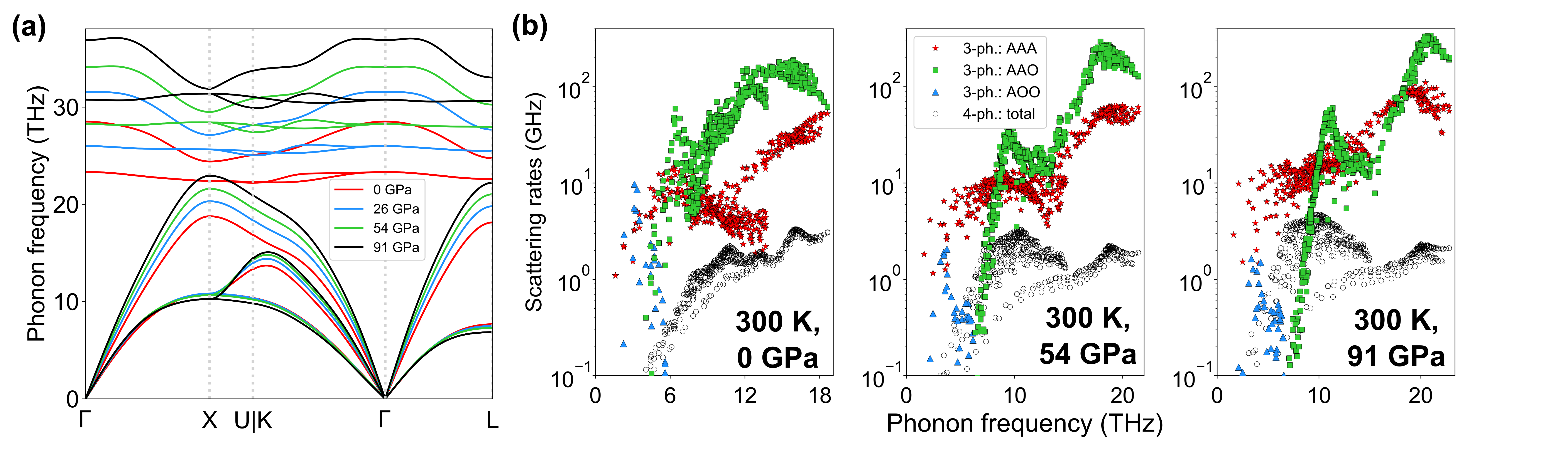}
\end{center}
\caption{\textbf{(a)} Phonon dispersions for SiC at 300 K and different pressures. \textbf{(b)} Process-wise separated three-phonon and total four-phonon scattering rates of SiC at 300 K and different pressures.}	\label{SI:fig_6}
\end{figure*}

\begin{figure*}[!ht]
\begin{center}
\includegraphics*[scale=0.275]{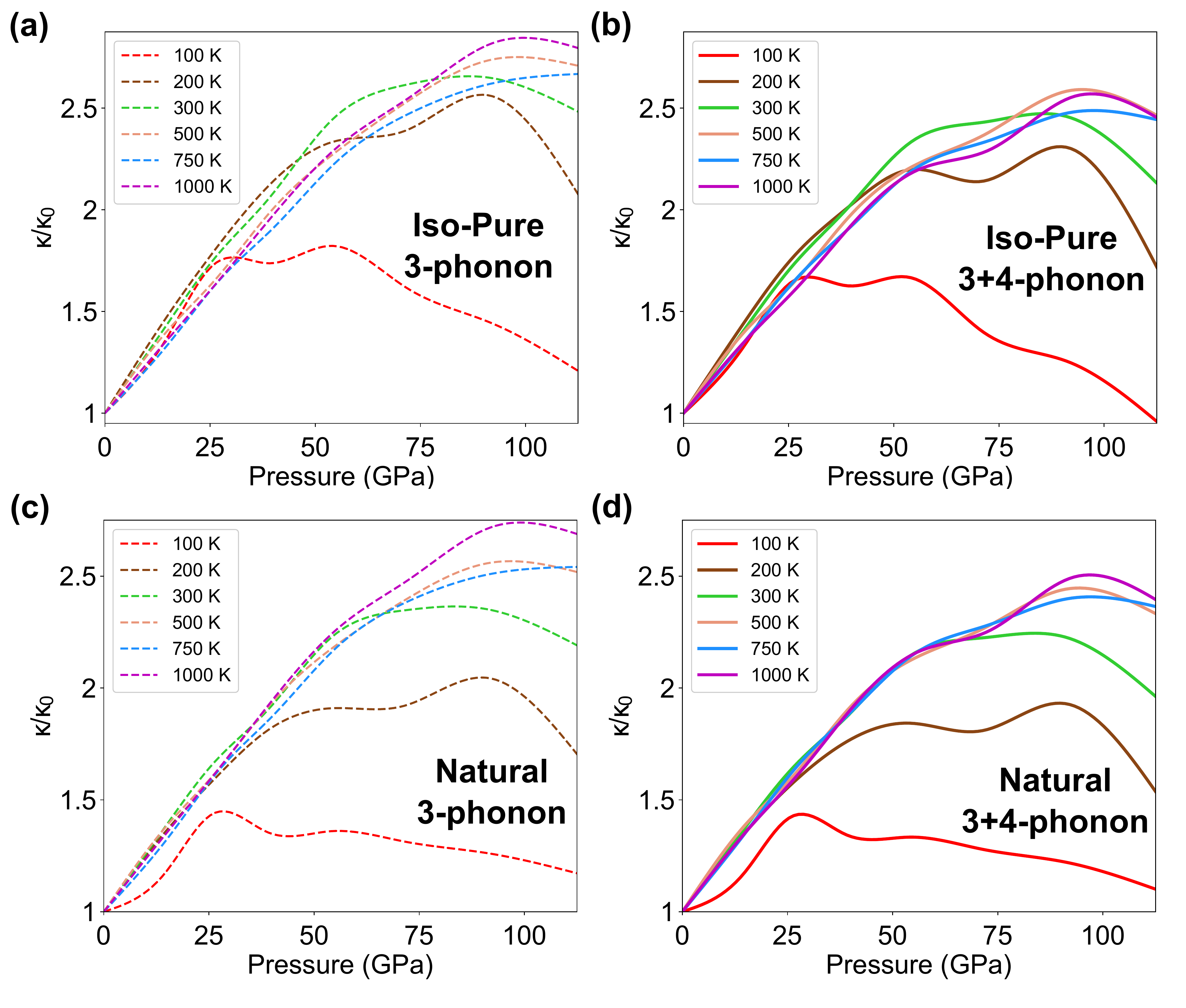}
\end{center}
\caption{Pressure dependence of the calculated thermal conductivity of isotopically enriched and naturally occurring cubic SiC scaled by its zero pressure values for different temperatures.}	\label{SI:fig_7}
\end{figure*}

\begin{figure*}[!ht]
\begin{center}
\includegraphics*[scale=0.5]{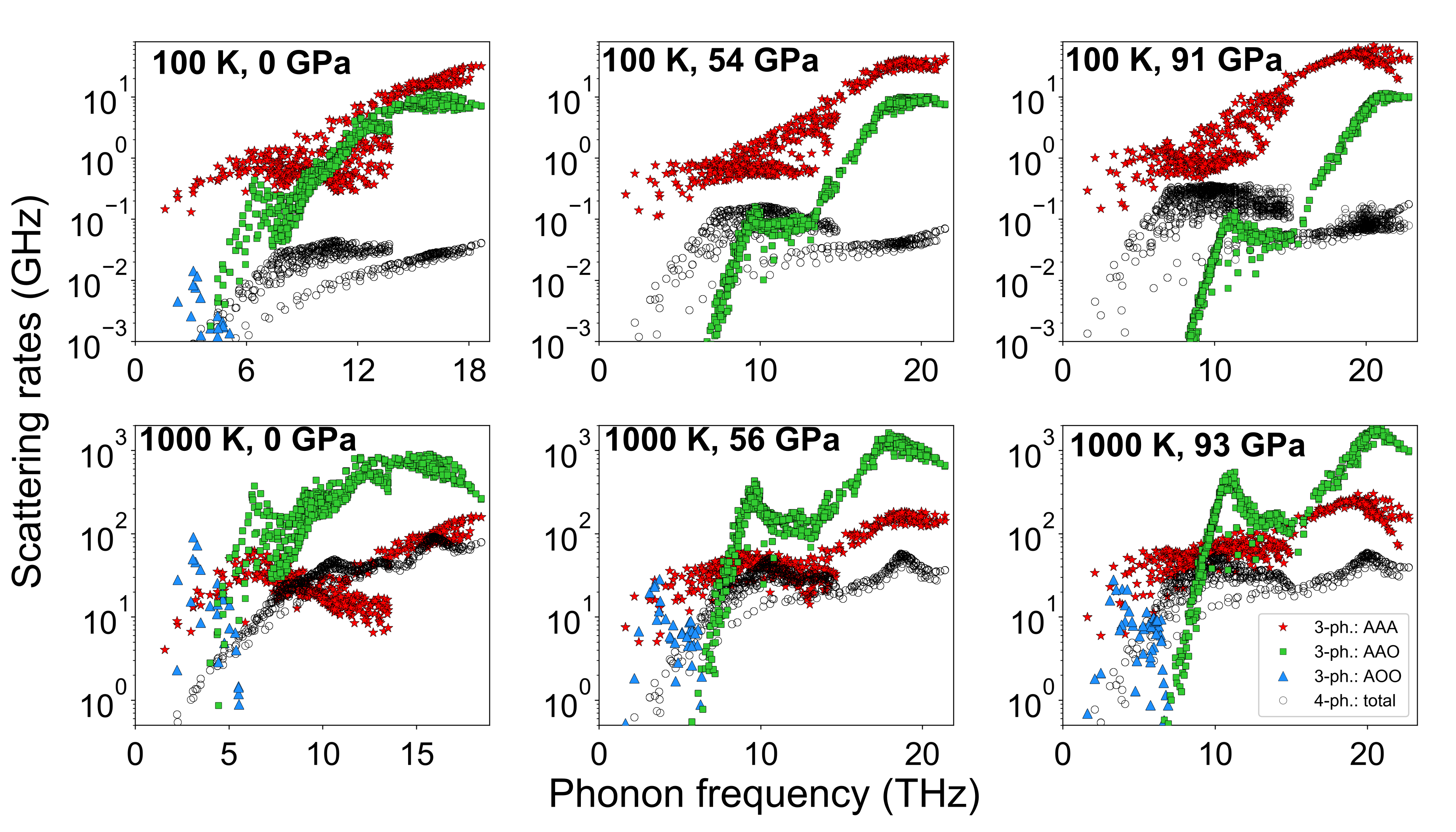}
\end{center}
\caption{Process-wise three-phonon and total four-phonon scattering rates for SiC at $T$ = 100 K and $T$ = 1000 K, and different pressures.}	\label{SI:fig_8}
\end{figure*}

\begin{figure*}[!ht]
\begin{center}
\includegraphics*[scale=0.4]{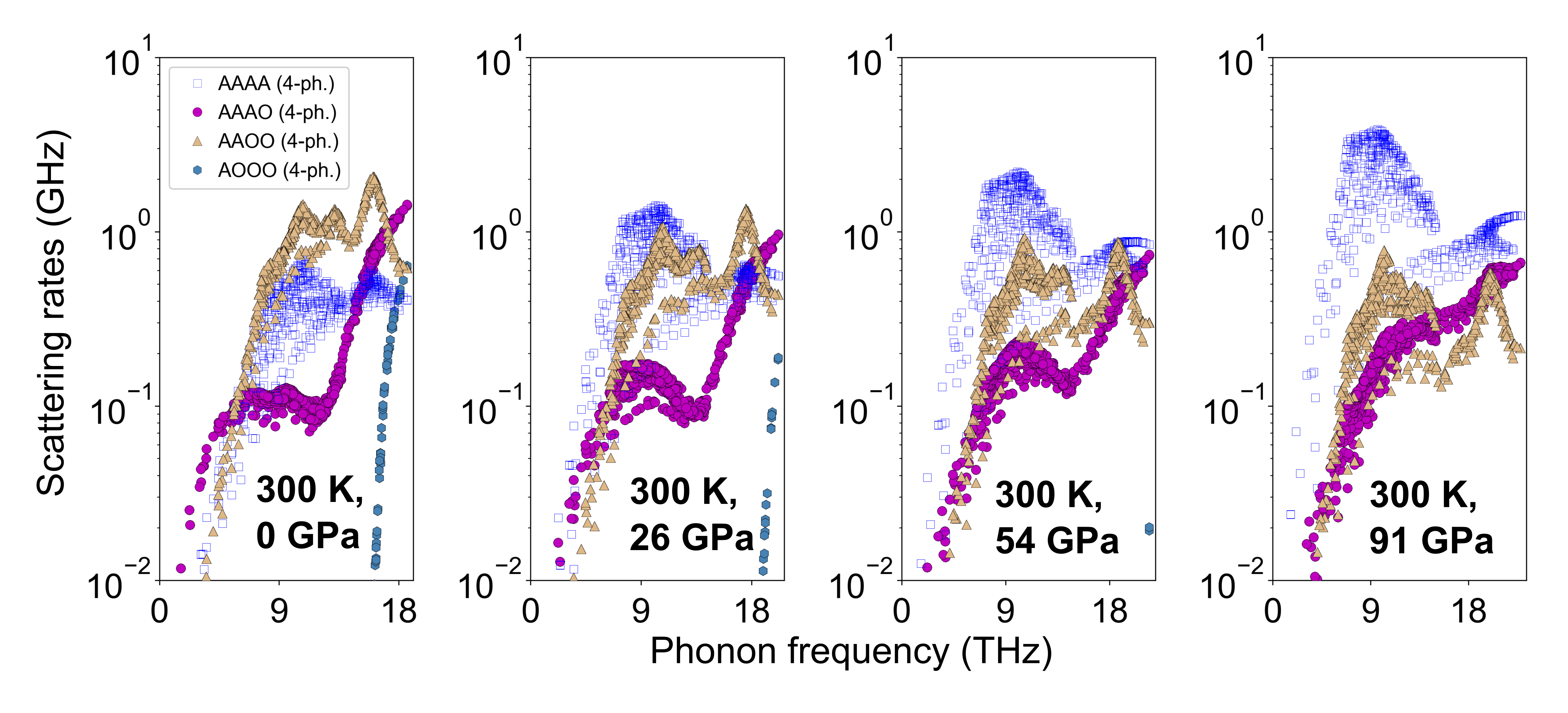}
\end{center}
\caption{Process-wise four-phonon scattering rates for SiC at $T$ = 300 K and different pressures.  \emph{AAAA} scattering rates increase with increasing $P$ and dominate at high $P$, qualitatively similar to the behavior in BP.}	\label{SI:fig_9}
\end{figure*}

The results for the pressure and temperature-dependent phonon dispersions, three-phonon and four-phonon scattering rates, and $\kappa$ for SiC are presented in this section. Figure~\ref{SI:fig_6} (a) shows the stiffening of the longitudinal acoustic and all optic phonon branches, and the weak softening of the transverse acoustic phonon branch of SiC with pressure. Figure~\ref{SI:fig_6} (b) shows the evolution of the process-wise separated three-phonon and total four-phonon scattering rates with pressure at room temperature. The two plots show the striking similarity of the trends found for SiC and for BP (in the main text). Similarly, application of hydrostatic pressure causes a peak and drop in $\kappa$ in SiC (Fig.~\ref{SI:fig_7}) as in BP (in the main text). Furthermore, the observed temperature- and pressure-dependence of the competition between \emph{AAA} and \emph{AAO} scattering rates in BP is also seen in SiC (Fig.~\ref{SI:fig_8}). Finally, the dominant all-acoustic \emph{AAAA} four-phonon scattering channel (see Fig.~\ref{SI:fig_9}) drives the reduction in $\kappa$ of SiC at high $P$, also similar to BP.\\
\end{document}